# Heterogeneous ferroelectricity and conductivity of oxidized $BaTiO_3$ crystals: the role of nanoscale phase segregation in the surface region

Christian Rodenbücher[1,*], Gustav Bihlmayer[2], Susan Trolier-McKinstry[3], Jacek Szade[4,†], Franciszek Krok[5], Kristof Szot[4]

[1]*Institute of Energy Technologies (IET-4), Forschungszentrum Jülich GmbH, 52425 Jülich, Germany*
[2]*Peter Grünberg Institute (PGI-1), Forschungszentrum Jülich GmbH, 52425 Jülich, Germany*
[3]*Department of Materials Science and Engineering and Materials Research Institute, The Pennsylvania State University, University Park, Pennsylvania 16802, USA*
[4]*Institute of Physics, University of Silesia, 41-500 Chorzów, Poland*
[5]*Marian Smoluchowski Institute of Physics, Jagiellonian University, 30-348 Kraków, Poland*

We investigate the effect of thermal oxidation on $BaTiO_3$ single crystals. Our results reveal that, even at moderate temperatures of up to 1000 °C, complex segregation mechanisms occur involving the movement of Ba-rich compounds to the upper surface, where they form inhomogeneously distributed BaO nanocrystals. As a result, deeper regions of the surface layer become depleted in Ba and become $TiO_2$-rich. A sandwich-like structure evolves in the surface layer, exhibiting a measurable electromotive force and self-polarization. This, in turn, screens the polarization and diminishes the global ferroelectric response. The conductivity is also strongly influenced by the irreversible segregation effects in the surface region. An as-received crystal can be transformed into a metallic state, while oxidation induces semiconducting properties and prevents a return to the metallic state upon subsequent reduction. Nanoscale analysis demonstrates that the conductivity is channeled to filaments forming along dislocations, whose local chemical composition changes upon reduction and oxidation even at moderate temperatures due to preferential Ba and O pipe diffusion thus determining the electric behavior of the whole sample.

## I Introduction

Although the terms "surface", "surface layer", "surface region", and "near-surface area" of perovskites such as $BaTiO_3$ (BTO) are often used in the same context, they can have completely different meanings in surface physics, defect chemistry, or materials science. From the perspective of surface physics, only the last few monolayers of the crystal are of relevance for changes in the crystallographic and electronic structure related to reconstruction, rumpling, or relaxation [1]. In defect chemistry describing the oxygen exchange in terms of Schottky disorder, the surface region is assigned to a space-charge zone with a thickness of a few dozen nanometers [2,3]. However, both approaches have drawbacks. If the thermal treatment of an oxide leads to significant crystallographic and chemical changes, one cannot only consider the role of the surface polarity as it is typically done in surface physics, e.g., by Tasker's criterion/classification [1] predicting stable configurations of surface atoms. When new phases form in the surface layer, the most appropriate approach to describe the underlying driving forces is applying thermodynamics. This approach can provide an initial prediction of the potential changes in chemical composition that may accompany oxidation and reduction processes in the surface layer, as demonstrated for BTO ceramics at temperatures above 1000 °C [4]. However, such a thermodynamic analysis might be oversimplified as it assumes only a statistical distribution of defects [5,6]. In reality, the surface layer of real ternary oxides can exhibit high concentrations of extended defects such as dislocations [7,8]. Hence, crystallographic shearing mechanisms and preferential diffusion along of dislocations should also be considered in order to understand the evolution of new phases in the surface layer upon thermal treatment of $ABO_3$-type oxides, especially at moderate temperatures ($T$ < 1000 °C) [9–13]. In consequence, micro-heterogeneities with an excess of AO or $BO_2$ compounds form in the surface region as observed for $SrTiO_3$ (STO) [14]. This effect manifests as nanoscale crystallites, which typically have a cuboid, prismatic or nanowire morphology and are often found near the exits of dislocations [8,15]. However, dislocation-related segregation effects are rarely considered in thermodynamic analyses and the literature on this topic is quite scarce [16–19].

* c.rodenbuecher@fz-juelich.de
† deceased

In this work, we investigate the influence of thermal oxidation on BTO single crystals. We analyze how changes in structure and composition in the surface layer influence the global properties of the crystal. The investigation presented here builds on our previous studies of STO and BTO under reducing conditions, in which we demonstrated that the insulator-metal transition induced by redox processes is linked to the dislocation-rich surface region, which features a complex 3D network of dislocations [20–22]. In selecting single crystals, we aimed not only to avoid the influence of grain boundaries and triple junctions, which are typical features of ceramics, but also to focus our investigation on surface-layer properties and the "skin" effect in BTO, a phenomenon primarily studied in single crystals. We limit our investigations to a temperature regime of 1000 °C, in which localized effects such as pipe diffusion can be expected. It should be noted that although the conventional sintering of ceramics is performed at temperatures well above 1000 °C and can thus be described by point defect thermodynamics [23], the sintering temperature of oxide ceramics can be significantly reduced by combined methods, such as two-step sintering, spark plasma sintering, microwave sintering, high-pressure sintering, flash sintering, or cold sintering [24]. To ensure compatibility of our results with thermodynamic descriptions and analyses based on first principles, we have selected oxidation and reduction times, which are commonly considered sufficient to achieve thermodynamic equilibrium. We employed crystals produced by top-seeded solution growth (TSSG), which are often regarded as ideal in the literature [25]. However, they still exhibit imperfections both in chemical composition and in the entire spectrum of one-, two-, and three-dimensional defects, which can be of high relevance for oxidation-induced processes.

## II Methods

Single crystals produced by top-seeded solution growth (TSSG) are employed (Mateck, Germany). The crystals with a thickness of 0.5 mm are epi-polished and (100) surface oriented. According to X-ray diffraction measurements the samples are $BaTiO_3$ single phase. A nearly stoichiometric Ba:Ti ratio of 0.5043:0.4947 is found by energy-dispersive X-ray spectroscopy with ZAF-correction on a representative sample. Before conducting experiments, the as-received samples are cleaned with methanol in an ultrasonic bath.

The electronic structure and stoichiometry of the surface are studied using an X-ray photoelectron spectrometer PHI 5800 (Physical Electronics, USA) with a monochromatic Al Kα X-ray source (1486.6 eV, 200–250 W) and a microfocus (300 × 700 μm$^2$). The pass energy (PE) is 23.5 eV, and the energy step 0.05 eV. Surface charging is compensated by a flood gun neutralizer. Data analysis and simulation are performed using the software MultiPak.

Calculations based on density functional theory (DFT) are performed within the local density approximation [26] using the full-potential linearized augmented plane-wave method [27] as realized in the FLEUR code [28]. The muffin-tin radii are chosen to be 2.3, 1.9, and 1.5 a.u. for the Ba, Ti, and O atoms, respectively. The basis function cutoff is 4.12 (a.u.)$^{-1}$ and the Brillouin zone is sampled with 8×4×1 k-points. All structures are relaxed with a convergence criterion of 0.05 eV/Å for the forces.

Time of flight secondary ion mass spectrometry (SIMS) is employed to obtain the chemical composition of the surface in static SIMS mode and in deeper regions of the surface layer by recording depth profiles in dynamic SIMS mode (TOF–SIMS 5, ION-TOF GmbH, Germany).

Investigations of the ferroelectric properties are performed by applying a voltage to the crystal in capacitor geometry while recording the current. Circular electrodes with a diameter of 1.5 mm and a thickness of 0.1 mm are deposited on the top and bottom side of the samples using Pt paste. The ferroelectric hysteresis loop is measured by the triangular voltage sweep method cycling the voltage between −400 and +400 V at a frequency of 5 mHz. The capacity as function of temperature is investigated by applying an AC voltage ($V_{ac}$ = 100 µV, $f$ = 25 kHz).

The electrical properties during thermal oxidation and reduction are investigated using a dedicated resistance measurement system as described in detail in previous publication [29]. Measurements are performed in a four-electrode configuration with a low AC polarization voltage (typically 4 mV). Thermal oxidation is realized by introducing pure oxygen in an externally heated UHV chamber at temperatures up to 1000 °C. The chamber is equipped with a calibrated mass spectrometer (E-Vision (MKS Instruments, USA) allowing for the investigation of effusion from the sample during thermal reduction.

For investigations of the topography, local conductivity, and piezoelectricity on the nanoscale an atomic force microscope (AFM) equipped with Pt-coated cantilevers is employed in contact and tapping-mode (JSPM 4210, JEOL, Akishima, Japan). Using a home-built external current-to-voltage converter, a current sensitivity in local-conductivity atomic force microscopy (LCAFM) down to the fA range is achieved.

## III Results

### Increase in the concentration of BaO in the surface layer of BTO upon oxidation

The analysis starts with an investigation of the influence of the electronic structure of the BTO surface layer upon oxidation. *Ex situ* X-ray photoelectron spectroscopy (XPS) with an information depth of 4 to 6 nm is performed for the as-received stoichiometric reference sample and for a sample oxidized at 500 °C under 200 mbar $O_2$ and subsequently cooled to room temperature (RT) rapidly. As the sample is exposed to ambient conditions for transfer to the XPS chamber after oxidation and cooling, an adsorption of contaminants such as CO, $CO_2$, OH, and $H_2O$ can be expected [22,30–32]. However, this method still allows for an investigation of oxidation-induced surface segregation, as was confirmed previously [33,34]. Figure 1a shows the spectrum of the Ba3d core line of the oxidized crystal in comparison to the as-received crystal. The deconvolution of the difference spectrum indicates that in addition to the main peak of the reference crystal at a binding energy $BE_1$ = 779.1 eV, two peaks with a higher binding energy $BE_2$ = 780.6 eV and $BE_3$ = 781.6 eV are present after oxidation. This indicates that the oxidation process leads to enrichment of the BaO concentration in the surface layer and a modification of the nature of its chemical bonding.

By comparing XPS spectra measured at different take-off angles (Figure 1b), one can observe that the intensity of the new oxidation-induced peaks is higher for a smaller take-off angle, revealing that they are present on the very surface. The valences of Ti in the surface layer do not change during oxidation (electron configuration $d^0$) and no occupied states can be identified close to the Fermi level (Figure 1c). However, oxidation increases the density of states (DOS) of the valence band (VB), particularly near its top (see difference spectra as blue and green curves in Figure 1c). This suggests that BaO segregation leads to the evolution of an additional state, which is potentially related to hybridization between Ba *s* and O *p* states and contributes to the total DOS of the oxidized crystal in the VB region.

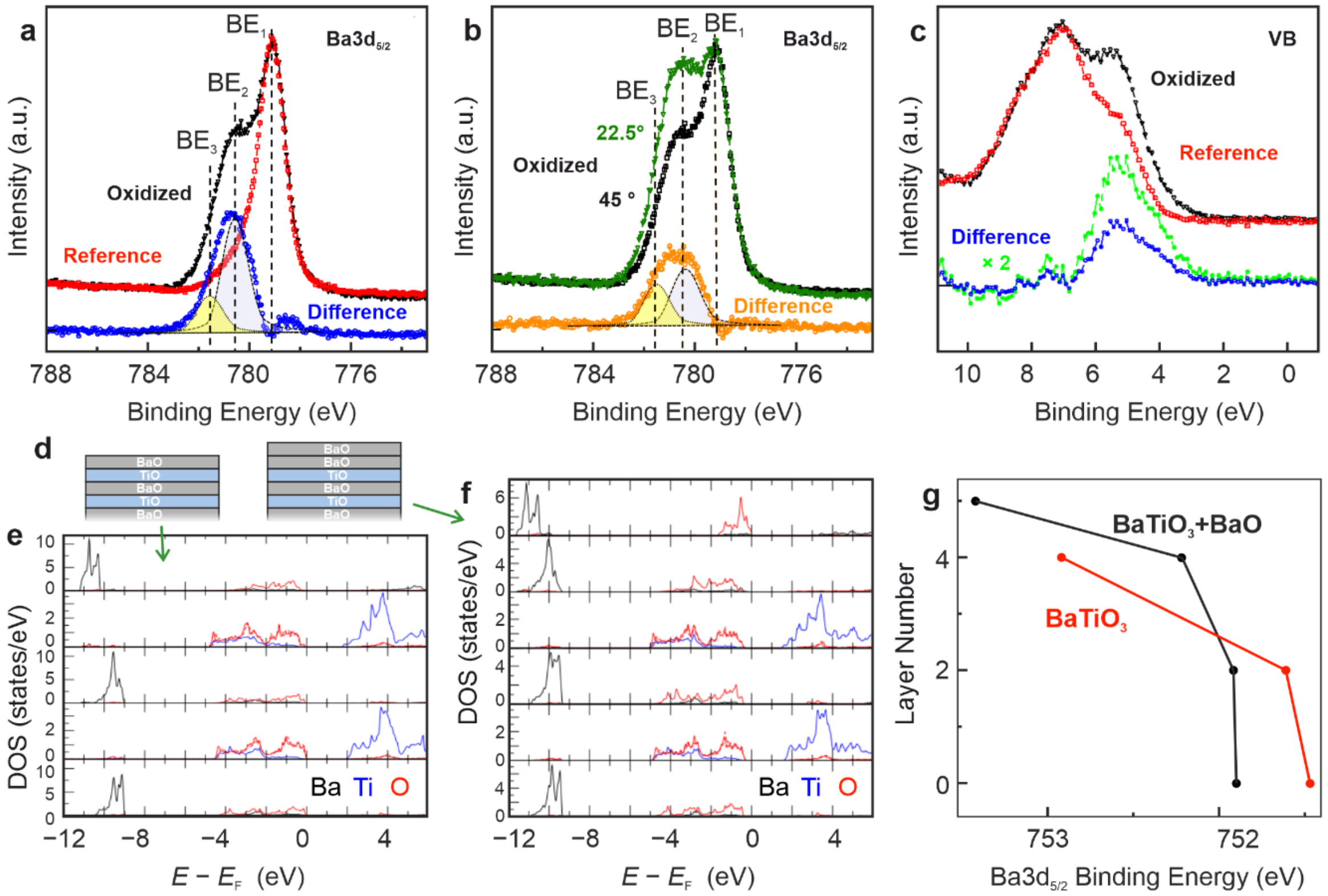

**Figure 1** Investigation of the influence of oxidation on the electronic structure. XPS analysis: a) $Ba3d_{5/2}$ core line before and after oxidation, b) oxidized $Ba3d_{5/2}$ core line at different take-off angles, c) valence band (VB) spectrum before and after oxidation; DFT simulation: d) setup of layers for DFT simulations of the surface and simulated DOS for the reference e) and the oxidized BaO-rich case f) calculated binding energy for the different layers.

## DFT calculation of the electronic structure of BaO-covered BTO

In order to give more evidence for the appearance of additional BaO layers on the surface upon oxidation, we perform density functional theory calculations for films of BaO terminated out-of-plane polarized BTO and its BaO-covered counterpart as illustrated in Figure 1d (note that only the upper half of the symmetric simulation cell is depicted). To stabilize the polarization, we use a 1×2×4.5 supercell with one 1×1×4.5 subblock exhibiting positive polarization and the other exhibiting negative polarization, similar to the setup introduced by Dionot et al. [35].

Comparing the DOS of a structure with the additional BaO layer (Figure 1f) to the one without (Figure 1e), additional O *p*-states from the extra layer can be identified at the Fermi level (that is, in the calculations, always at the top of the valence band). Compared to the states from the inner layers, they are about 0.5 eV higher in energy and have a width of slightly less than 2 eV. These additional states correspond to the features observed in the XPS difference spectra between oxidized and stoichiometric samples (Figure 1c). In these DOS plots, the shift of the Ba 5p states in the surface layers is also shown. Additionally, one can analyze the simulated Ba $3d_{5/2}$ core levels for the two situations, as shown in Figure 1g. We observe a strong surface core-level shift to higher binding energies, approximately 1.3 eV in both cases. The core levels of the additional BaO overlayer appear at higher binding energies due to the changed Fermi level (see Figure 1c). This is in line with the spectra shown in Figure 1a where an even larger surface core-level shift can be seen as well as a component at still higher binding energies for the oxidized surface, thus serving as further indication for the segregation of BaO during oxidation.

**SIMS investigation of oxidized BTO crystals**

The chemical composition of the surface layer is further analyzed by SIMS analysis in ***i.*** static mode, which is sensitive to the topmost surface, and in ***ii,*** dynamic mode, which gives access to the deeper parts of the surface layer. To avoid a possible influence of batch-to-batch variation, one BTO crystal piece is cut into four samples. One is taken as reference and three are annealed in 200 mbar oxygen for 1 h at temperatures of 800, 900, and 1000 °C, respectively. Figure 2a shows the Ba/Ti ratios of the surface of the oxidized samples normalized to the reference sample as obtained by static SIMS. The results reveal a systematic increase in Ba concentration on the surface with annealing temperature. To improve the statistical accuracy of the results, SIMS measurements are conducted in different regions of the annealed samples (see red data points in Figure 2a). The large variation in the Ba/Ti ratios (integrated over an area of approximately 4,000 μm² each) indicates significant lateral heterogeneity in the distribution of barium in the oxidized crystal plane.

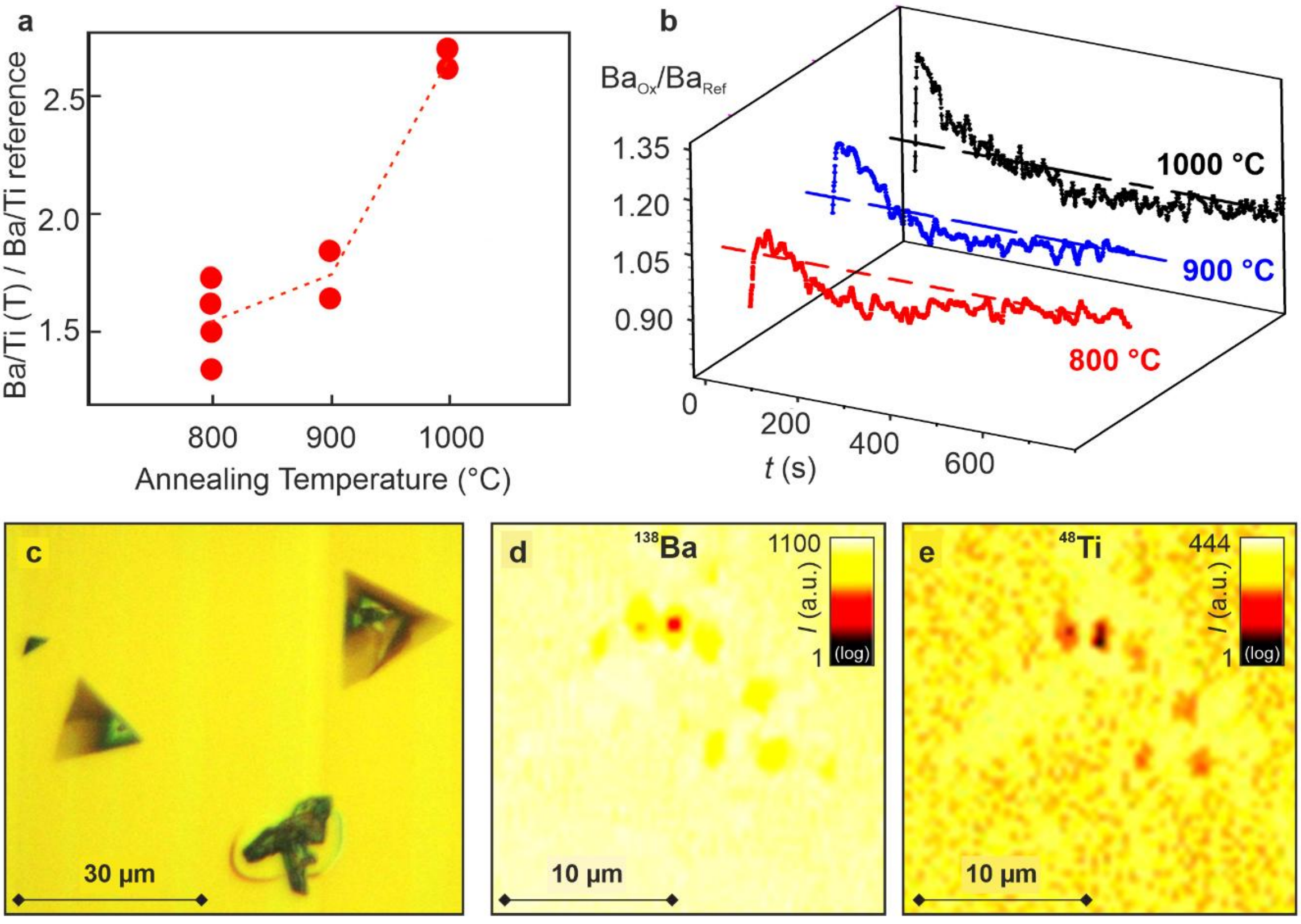

**Figure 2** SIMS analysis of oxidized BTO samples. a) Normalized Ba/Ti ratio of the surface as function of the annealing temperature obtained by static SIMS, b) dynamic SIMS depth profiles of samples oxidized at different temperatures (the investigated depth corresponds to several hundred nanometers), c) optical microscopy of a BTO crystal oxidized at 1000 °C, and d) $^{138}$Ba and e) $^{48}$Ti SIMS mapping.

Depth profiles of Ba (normalized to the Ba concentration of the reference sample) are obtained by dynamic SIMS for different oxidation temperatures (Figure 2b). Increasing the annealing temperature results in continuous enrichment of Ba in the upper part of the surface region. This Ba enrichment at the surface is accompanied by Ba depletion in the deeper parts of the surface region, which indicates that the region close to the surface becomes BaO-rich, while the deeper parts are enriched in $TiO_2$. Given that the stoichiometry range for a BaO surplus and $TiO_2$ deficit in BTO is low and similar to that for STO [36,37], it can be expected that the relatively large changes in chemical composition lead to a structural transformation involving Ruddlesden-Popper (R-P) and Magnéli phases, as observed, e.g., by *in situ* X-ray diffraction (XRD) [33,34]. Also in BTO ceramics exposed to oxidizing conditions at temperatures above 1200 °C, segregation of BaO to the surface has been found [23]. Our depth profiles demonstrate that in a single crystal, this segregation process of Ba or BaO is not only limited to the uppermost surface layer, but involves a much deeper region, extending several hundred nanometers.

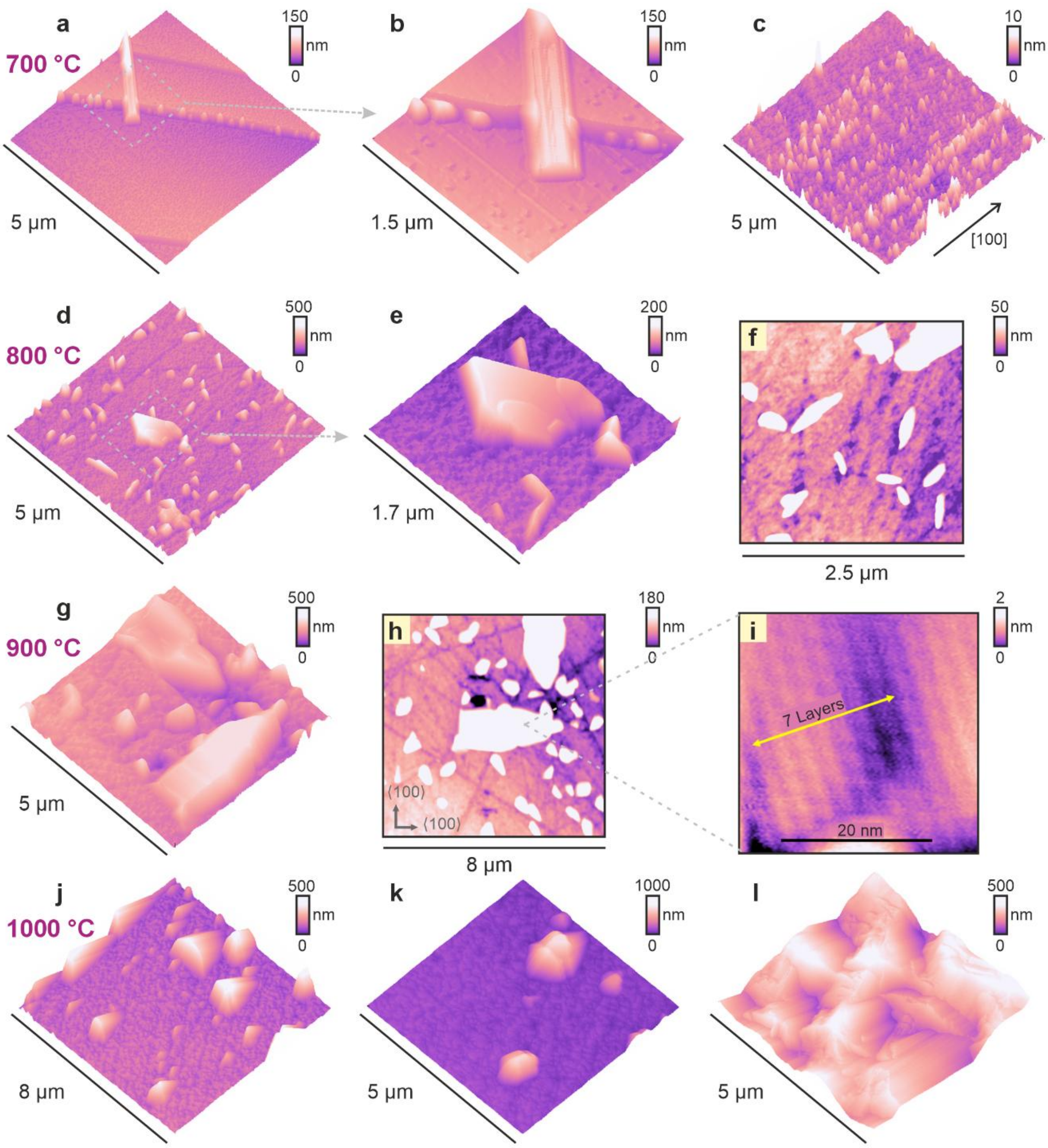

**Figure 3** Topography of BTO crystals after oxidation at a) 700, b) 800, c) 900, and d) 1000 °C investigated by tapping-mode AFM.

The tendency of growing crystals to increase in volume is maintained when they are annealed at 900 and 1000 °C (Figure 3g–l). Topographical AFM images of oxidized crystals at 900 and 1000 °C show areas in which the crystallites cover the entire surface. Regions in which the initial surface is visible between the crystallites are less common. In these regions, the thermally etched bands of dislocations can be more clearly identified than on the surface after thermal treatment at 800 °C (Figure 3h). In order to obtain information about the crystallographic structure of the crystallites, their surface was mapped with high resolution (Figure 3i). A regular arrangement of atomic layers with a periodicity of 2.8 nm is observed. This distance is typical of R-P phases ($BaO(BaTiO_3)_3$) [33,34,38,39].

**Ferroelectric characteristics of oxidized BTO**

In the following, we investigate the impact of oxidation of BTO single crystal on its ferroelectric properties. While this has been done extensively for ceramics annealed at temperatures above 1000 °C (e.g., [34,40]), we here limit the oxidation temperature to 1000 °C. At this temperature only the formation of point defects is expected according to defect chemistry [3]. Analyzing the modifications of ferroelectric behavior induced by oxidation requires confirming that the crystals used in our experiment initially exhibit typical properties of TSSG grown crystals [41,42]. Therefore, we measure the ferroelectric loop at very low frequencies (5 mHz) using the triangular voltage sweep (TVS) method to determine the polarization and to investigate leakage currents. Note that we use Pt paste electrodes with large grains which allow for oxygen exchange between the sample and the ambient. Figure 4a demonstrates that the ferroelectric loop exhibits a perfect shape with a remanent polarization about 19 µC/cm², slightly smaller than the spontaneous polarization $P_S$ reported in the literature for BTO single crystals (26 µC/cm²) [43,44] and ceramics (20–25 µC/cm²) [45]. A possible explanation for the observed deviation could be that the surface region of the polished single crystals contains a large concentration of dislocations, which can strongly reduce the switchable polarization [46].

The measured hysteresis loop of the oxidized BTO crystal (Figure 4b) differs significantly from the as-received case. Let us ask, which ferroelectric response could be obtained from the oxidized BTO crystal consisting of a stoichiometric interior, which is covered by layers of different composition according to our XPS and SIMS analysis? Note that this scrutiny of the influence of the surface layer of oxidized crystals on their ferroelectric properties follows the traditional analysis of the "skin" effect in BTO [44]. The first conclusion that comes to mind when analyzing the dielectric or ferroelectric response of a system with different chemical compositions in its surface area and interior is a change in the electric field distribution in these areas due to their differing electrical conductivity and dielectric permittivity. Therefore, it is expected that, e.g., the coercive field for both regions will not be the same, and in the end, the nonlinear part of the hysteresis loop (connected with the reversal of spontaneous polarization) of the whole system (here two layers with bulk) will be "blurred". Indeed, Figure 4b reveals that the polarization has been reduced to a small fraction of that determined for the reference crystal. This indicates that the polarization is strongly screened by the modified surface layer of the oxidized crystal. Note that polarization and the dielectric properties of the reference and oxidized crystals are measured using crystals prepared from the same large substrate.

In Figure 4c, we investigate the capacity as function of temperature during heating across the ferroelectric/paraelectric transition. Note that we intentionally plot the capacitance and not the dielectric constant (permittivity) here, as the presence of a near-surface region with a high concentration of dislocations can influence the dielectric properties in a way which is not fully understood yet. Hence the system surface layer/crystal interior/surface layer should be understood as a Maxwell-Wagner capacitor. For the stoichiometric crystal (blue curve in Figure 4c), a sharp increase of the capacity at a temperature of $T_C = 132$ °C is observed, which is a typical Curie temperature for melt grown crystals [47]. After oxidation at 1000 °C, only small but clear differences in the capacity-temperature dependence can be observed (red curve in Figure 4c). The phase transition can still clearly be identified, but the increase in capacitance becomes less sharp. Since the surface region of the oxidized crystal is much thinner (~100 nm) than the bulk region (~500 µm), the displacement current is dominated by the interior of the crystal, which retains its original composition. Additional peaks in $C(T)$ below and above the original $T_C$ temperature of 132 °C are also visible. They could be related to the segregation of BaO-rich compounds in the upper surface region accompanied by a relative $TiO_2$ enrichment in deeper regions as confirmed by SIMS depth profiles (Figure 2b). It is known from ceramics that the phase transition temperature is highly sensitive to the composition and decreases by a few degrees relative to the stoichiometric $T_C$ for BaO-rich structures and increases by a few degrees for $TiO_2$-rich BTO [48,49].

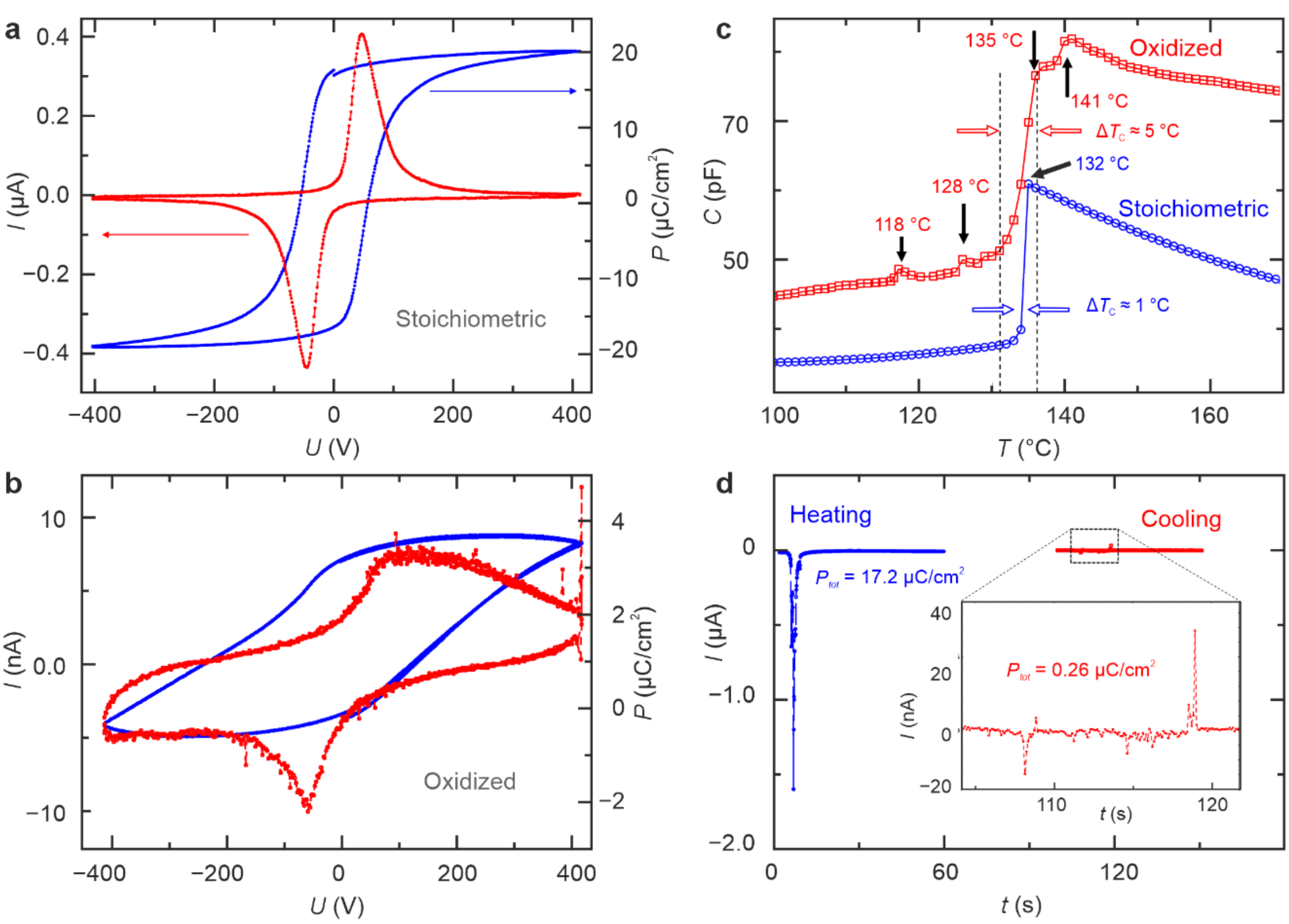

**Figure 4** Analysis of the ferroelectric properties. Ferroelectric current loop (red) measured at a frequency of 5 mHz and polarization (blue) for a) as-received and b) oxidized ($T_{ox} = 1000$ °C, $t_{ox} = 4.5$ h) BTO (the ferroelectric polarization loop has been generated by integrating the current curve); c) capacity measured with an alternating voltage ($V_{ac} = 100$ µV, $f = 25$ kHz) of an as-received (blue) and oxidized (red) BTO crystal during heating cycles between the ferroelectric and paraelectric phase; d) depolarization current of previously polarized stoichiometric BTO during heating (blue) and b) subsequent cooling (red) with a constant rate of $dT/dt \approx 0.5$ °C/s around $T_C$.

For completeness, we also verify the ferroelectricity of the stoichiometric sample by depolarizing a previously poled sample. Therefore, the sample is heated to 160 °C, an electric field much higher than the coercive field is applied, and the sample is cooled under field to room temperature. Subsequently, the sample is heated at zero external field from the ferroelectric to the paraelectric phase at a constant rate. The crystal is short-circuited by a virtually grounded I/V converter. As expected, measuring the depolarization current yields the amount of electronic charge released, which screens the polarization, to 17 μC/cm² (blue curve in Figure 4d). Upon subsequent cooling, the sample does not show any current above the noise level during the paraelectric-ferroelectric transition as it is completely depolarized (red curve in Figure 4d).

To better understand the influence of the surface region of the oxidized BTO crystal on the ferroelectric properties, we remove one side via mechanical polishing, as illustrated in Figure 5a, thereby creating an asymmetric configuration. This approach follows our previous investigation of the self-polarization effect in oxidized $PbTiO_3$ single crystals [50]. After removing a few dozen micrometers of one surface and depositing Pt paste electrodes, the current during repeated heating and cooling cycles is measured without external polarization. One can observe the depolarization and polarization current during both heating and cooling across $T_C$ (Figure 5c), indicating that the system is self-polarized. The polarization values ($P^+ \approx 4$ μC/cm², $P^- \approx 4$ μC/cm²) determined by integrating the depolarization curves are of same order of magnitude as the reference depolarization current of the poled stoichiometric crystal (Figure 4d). Note that after polishing off both surface layers and removing the edges, the self-polarization effect of the oxidized crystals vanishes completely, and we can only observe Barkhausen noise. This behavior suggests that in the surface region, an electrochemical cell with a potential difference (electromotive force, EMF) evolves during oxidation, related to the BaO segregation causing a concentration gradient of Ti with different valences similar to the case for $PbTiO_3$ [50]. This leads to the presence of a strong electric field in the surface region between the BaO-rich and $TiO_x$-rich layers. In consequence, a thin part of the crystal can self-polarize as illustrated in Figure 5a. Hence, for the asymmetric configuration of the polished oxidized sample, polarization and depolarization currents can be measured, which originate from the surface region.

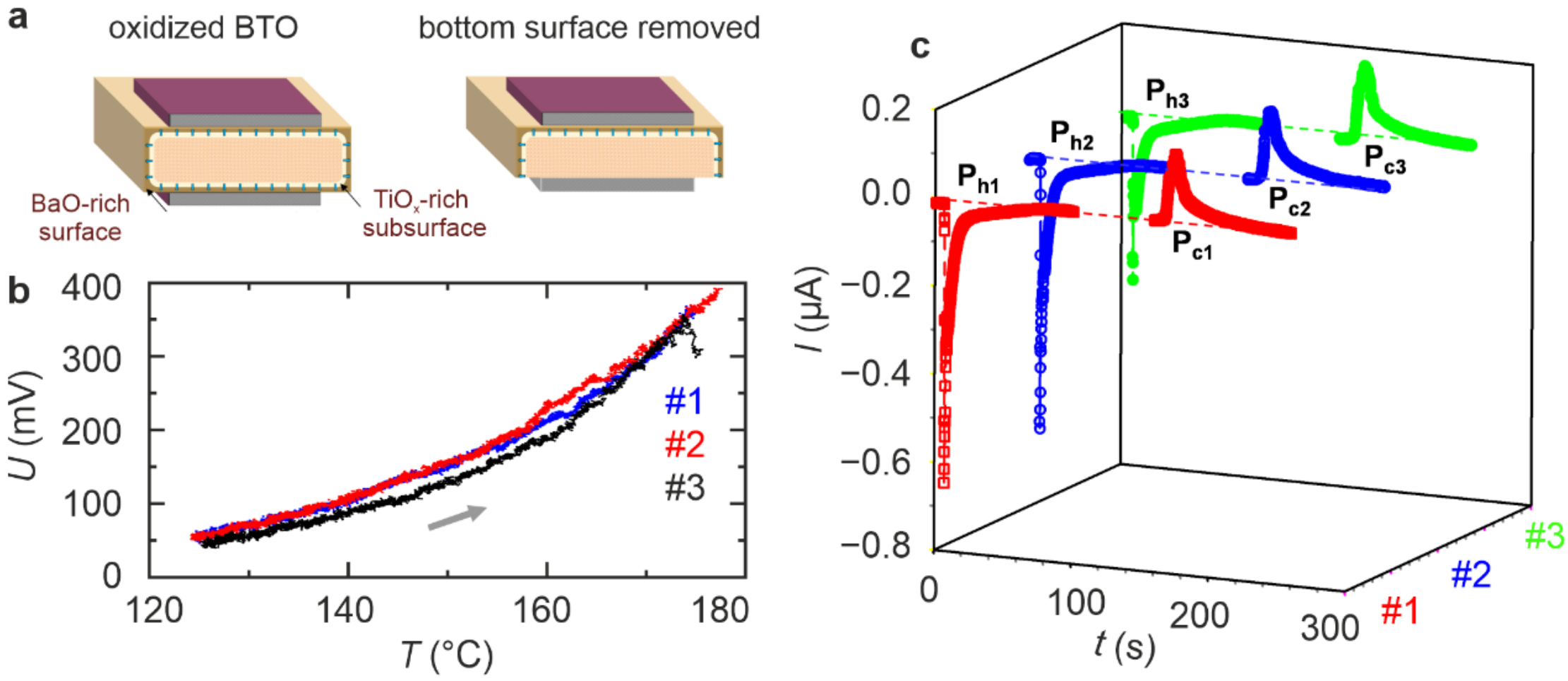

**Figure 5** Investigation of a one-side polished oxidized BTO crystal. a) Illustration of the mechanical removal of one surface and subsequent electrode deposition, b) measurement of the voltage (electromotive force, EMF) for the asymmetric configuration as a function of temperature in a series of successive heating cycles, c) depolarization- and polarization-current of asymmetric configuration obtained by repeated heating (h1 – h3) and cooling (c1-c3) of the crystal with a constant heating/cooling rate $dT/dt$ across $T_C$.

In order to confirm the existence of an internal electrical field in the surface region of the oxidized crystal, electromechanical measurements of the EMF are performed as function of temperature (Figure 5b). As the temperature increases, the EMF value also increases. The slight deviation from the linear dependence of the EMF on temperature, which should be observed for a cell with an electromotive force that fulfils Nernst's law [51] might be caused by changes in the cell's internal resistance related to an inhomogeneous distribution of the BaO-rich structures in the surface region as revealed by AFM and SIMS.

Piezo force microscopy (PFM) mapping is employed to obtain more information about the effects of inhomogeneities in the surface region on ferroelectric properties. A 4×4 µm region is mapped at different temperatures across the phase transition. The topography map in Figure 6a reveals that the surface roughness increased significantly compared to the epi-polished as-received surface upon oxidation at 1000 °C. The corresponding maps of the PFM out-of-plane amplitude measured between 110 and 150 °C reveal a very inhomogeneous spatial distribution, which could indicate that the transition between ferroelectric and paraelectric state occurs at different temperatures in different regions (Figure 6b-f). A progressive decrease of the integral piezoresponse of the analyzed region with increasing temperature can be observed. However, at 150 °C, the boundaries between the crystallites still exhibit residual piezoelectric activity. To assess the local variation of the Curie temperature on the oxidized BTO surface in more detail, we measure the PFM amplitude as a function of temperature at various randomly selected PFM tip locations. At a fixed tip position, only the piezo activity in a small area equal to the potential drop area close to the tip's apex influences the measured amplitude [52]. In this configuration, we can determine the onset of piezoelectricity on the nanoscale and define the local $T_C$ while cooling the sample from 180 to 80 °C (Figure 6g). Our PFM inspection at different positions reveals that, in many regions, the temperature of the step-like jump in piezoresponse shifts towards a lower temperature than the stoichiometric $T_C$, which is typical for a surplus of BaO [48]. However, some positions also show an increase of the local $T_C$, characteristic of regions with a higher $TiO_2$ concentration [49]. This behavior suggests that BaO segregation does not only occur out-of-plane but also in-plane. Additionally, it can be seen in Figure 6g that in positions with a higher local $T_C$, the PFM amplitude is slightly higher than in regions with a lower $T_C$.

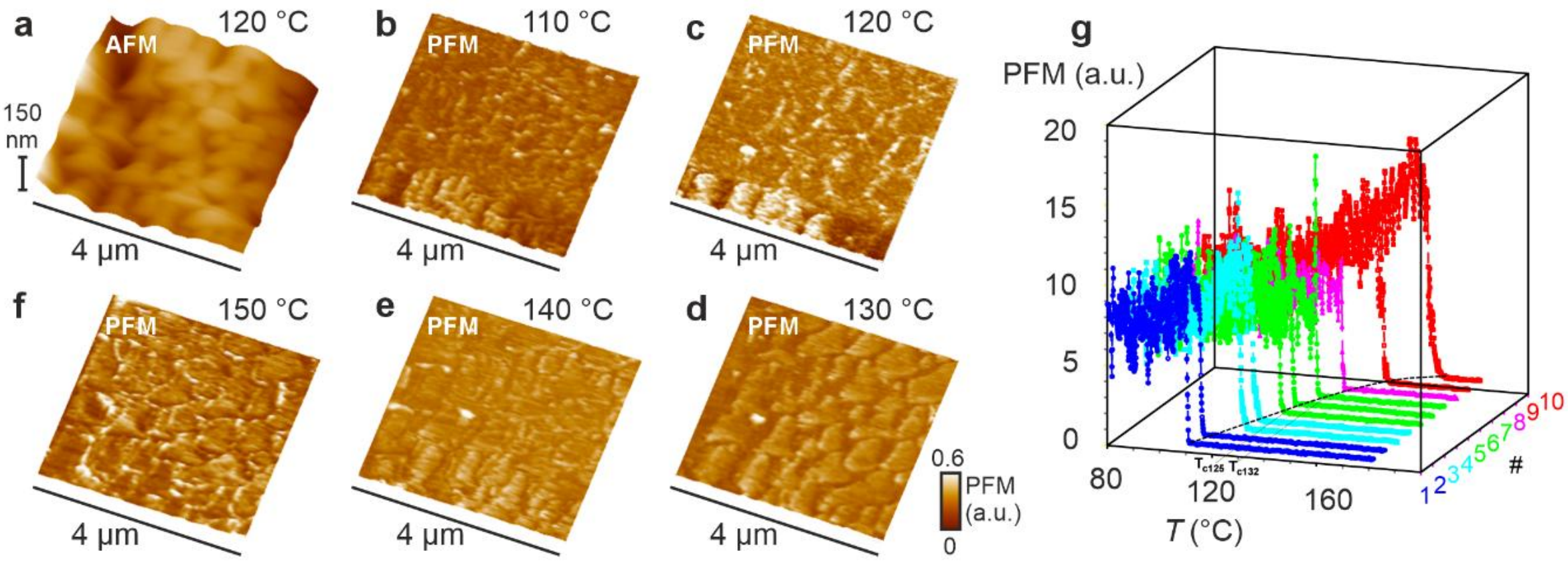

**Figure 6** Spatially resolved investigation of ferroelectric properties of an oxidized crystal. a) Topography (AFM) mapping and b-f corresponding maps of PFM amplitude (out-of-plane); g) local measurements of the PFM amplitude recorded during cooling of the crystals for different randomly selected positions of the AFM-tip.

**Defect chemistry**

In this section, we examine how the transformation and recrystallization of the surface region upon oxidation affects electrical transport phenomena. In other words, our goal is to investigate the role of the surface region in understanding point defect chemistry at low and high oxygen partial pressures. When discussing point defect chemistry, one should be aware that commercial BTO crystals contain impurities, which strongly influence the defect formation despite their small concentration. For example, Suzuki et al. found as dominant acceptors $Ag^{+}$ in the BaO sublattice and $Y^{3+}$ in the $TiO_2$ sublattice (~100 ppm), while the donor concentration in the $TiO_2$ sublattice ($Nb^{5+}$, $W^{6+}$, and $Mo^{6+}$) was lower (<60 ppm) [53]. Therefore, commercial BTO single crystals should be regarded as slightly acceptor-doped and the following defect chemistry equations should be considered:

$$O_O^x = V_O^{\bullet\bullet} + 2e' + ½O_2(g)$$

$$nil = e' + h^{\bullet}$$

$$nil = V_{Ba}'' + V_{Ti}'''' + A_C' + 3V_O^{\bullet\bullet} + D_C^{\bullet}$$

$$n = [e']; p = [h^{\bullet}]$$

$$n + 2[V_{Ba}''] + 4[V_{Ti}''''] + [A_C'] = p + 2[V_O^{\bullet\bullet}] + [D_C^{\bullet}]$$

$$for\ [A_C'] \gg [D_C^{\bullet}]$$

$$n + 2[V_{Ba}''] + 4[V_{Ti}''''] + [A_C'] = p + 2[V_O^{\bullet\bullet}];$$

*Index $_C$: refers to impurity atoms in the position of cations (Ba or Ti)*

An exemplar Brouwer diagram of acceptor-doped BTO at 1000 °C is illustrated in Figure 7a describing the defect concentrations as function of oxygen activity. At low oxygen activity, n-type conductivity prevails, while at high oxygen activity, BTO is a p-type semiconductor. This change in dominant charge carries directly influences conductivity. Hence, the best test for validating the point defect chemistry of BTO is the determination of electrical conductivity at high temperatures under different oxygen activities.

We measure the conductivity by a physical method, which relies on dosing pure oxygen in a UHV chamber, thus avoiding the need of gas mixtures [29,54]. The adjustment of the oxygen activity is performed by introducing a defined dose of oxygen into the vacuum chamber, thereby stabilizing the chamber pressure at a defined level for a constant pumping speed of the turbomolecular pump. Hence, the oxygen activity equals the total pressure. The accuracy in controlling the oxygen pressure for UHV and high vacuum is 1% per decade. For pressures higher than $10^{-2}$ mbar, the precision is 0.1%. This allows us to directly compare data obtained by surface-sensitive methods, which provide precise information about the electronic structure and chemical composition, with electrical transport data. Thus, we can trace the contributions and modifications to electrical transport arising from long-term oxidation of reduced BTO crystals, while accounting for important boundary conditions, including the physics and chemistry of the surface layer. Figure 7b depicts our data of the electrical conductivity of the BTO crystal as a function of oxygen activity at 1000 °C in comparison to literature data [55–62]. Our data corresponds well to the published data, confirming the validity of the applied physical method. Note that for calculating the conductivity, we here (incorrectly) assume a homogeneous distribution of charge carriers in the whole volume of the sample following the traditional approach of investigating point defect chemistry experimentally. However, we know from our surface-sensitive measurements that the dislocation-rich surface region strongly affects the electric transport of the whole sample by current channeling and nanoscale phase separation upon oxidation. Hence, we believe that calculating a global conductivity $\sigma$ for an inhomogeneous system is misleading and we will use the total resistance $R$ only in the following.

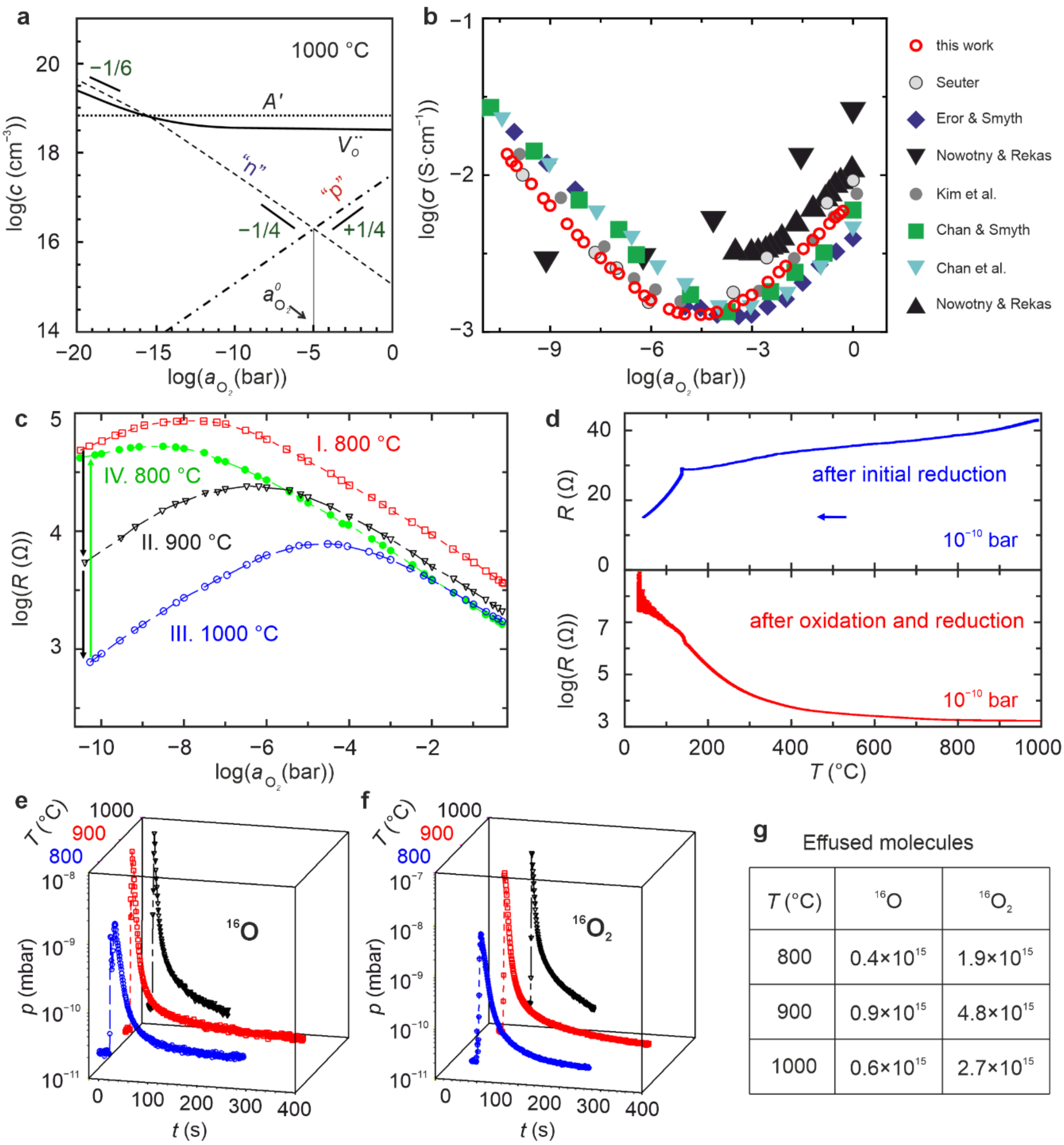

| $T$ (°C) | $^{16}O$ | $^{16}O_2$ |
|---|---|---|
| 800 | $0.4\times10^{15}$ | $1.9\times10^{15}$ |
| 900 | $0.9\times10^{15}$ | $4.8\times10^{15}$ |
| 1000 | $0.6\times10^{15}$ | $2.7\times10^{15}$ |

**Figure 7** Point defect chemistry investigations of BTO single crystals. a) Brouwer diagram of acceptor-doped BTO at 1000 °C adapted from Yoo et al. [55]; b) comparison of the conductivity as function of oxygen activity measured in this work with literature data compiled by Yoo et al. [55]; c) Total resistance as a function of oxygen activity, measured with our physical method; d) resistance as function of temperature measured in UHV conditions for initially reduced BTO (blue) and oxidized and subsequently reduced BTO (red); analysis of the effusion of e) atomic and f) molecular oxygen during thermal reduction in UHV of BTO crystals that were previously reduced and reoxidized; g) number of effused molecules calculated from the effusion data in e) and f).

Figure 7c illustrates the measurement of total sample resistance as function of oxygen pressure in the chamber at different temperatures. We perform measurements of the sample resistance at different temperatures of 800, 900, and 1000 °C while changing the oxygen activity by dosing pure oxygen in the vacuum chamber. The slope of the branches in the “p” and “n” regions of the double-logarithmic *R* vs. $a_{O_2}$ plot is at all temperatures in the range of 0.22 and 0.23, and thus a bit smaller than the expected value of 0.25. From the position of the intrinsic resistance maximum, we can calculate the activation energy corresponding to the formation enthalpy of an electron–hole pair [3], $E_i$ = 1.47 eV. Performing the same calculation for the “p” and “n” regions, we obtain the corresponding values of activation energies: $E_p$ = 0.5 eV and $E_n$ = 2.46 eV.

Since reduction or oxidation transforms the chemical composition of the surface region, as evidenced by the growth of BaO-rich crystallites on the surface during oxidation or the formation of low-Ti suboxides during reduction [22], it is necessary to verify the repeatability of the diagrams. To do so, we examine the possibility of reproducing the initial resistance isotherm at 800 °C after the pressure-resistance cycles at 900 and 1000 °C (which is used for the comparison with literature data in Figure 7b). After cooling the crystal from 1000 to 800 °C for 1 day under UHV conditions (green arrow in Figure 7c), the resistance returns to its original value, as in the first measurement. However, the resistances measured at higher oxygen activities do not match the initially measured ones but are significantly lower, indicating that irreversible processes on the surfaces have taken place during high-temperature oxidation and reduction (compare the green and red line in Figure 7c).

Irreversible processes are even more evident when analyzing the initial reduction step, i.e., when an as-received BTO crystal is annealed to high temperatures, here, 1000 °C under UHV conditions for the very first time. This step can directly be compared to effusion measurements, determining the amount of oxygen removed during reduction and to operando XPS studies of the composition and electronic structure of the surface layer [22]. When we compare the calculated concentration of oxygen vacancies in polycrystalline BTO at 1000 °C [55] with our effusion data obtained at the same temperature [22], we find that the amount of removed oxygen is 3.5 orders of magnitude smaller than the expected concentration of oxygen vacancies originating from unintentional acceptor doping introduced during crystal growth. Despite the significant modification of the surface region during reduction and oxidation, our *in situ* XPS analysis shows that, after removing the mentioned layer, the crystal volume remains unchanged after high-temperature treatment. Although only an extremely low concentration of oxygen vacancies is incorporated into the crystal by reduction, the total resistance drops to extremely small values around 40 Ω and further increases when cooling down the crystal under UHV conditions to room temperature (blue curve in Figure 7d). This indicates that a transition from insulating (I) into a metallic (M) state occurs, which is highly reproducible as we have observed it for all virgin (as-received) BTO and STO [54] crystals. As discussed in detail in our previous publication [22], we found evidence that only the preferential reduction of the three-dimensional network of dislocations in the surface region is responsible for this transition. When removing a few micrometers of the surface region by *in situ* abrasion, XPS measurements revealed that a surface created in this way does not exhibit metallic behavior, such as an occupied electronic state at the Fermi level or lower Ti oxidation states (3+ and 2+), but the XPS spectra are identical to those of the stoichiometric insulating crystal. The amount of effused oxygen, as determined by the reduction of the BTO crystal operando, does not allow us to calculate a global oxygen vacancy density in the crystal, because the initial reduction process is limited to the surface region resulting in a nonuniform distribution of oxygen vacancies in a reduced crystal, even though electrical measurements show that the crystal has reached a minimum global resistance [54,63]. However, such a metallic state, which is not expected from homogeneous point defect chemistry, can only be induced during the first UHV reduction of an as-received crystal. Once the crystal is exposed to oxidizing conditions at high temperatures, the metallic state vanishes. Even when the oxidized crystal is exposed to the same oxygen partial pressure and temperature, the global resistance remains orders of magnitude higher than in the metallic state and the resistance-temperature dependence indicates semiconducting properties (red curve in Figure 7d). As soon as the crystal is oxidized at high temperature, it behaves according to point defect chemistry, exhibiting the characteristic oxygen activity dependence of the resistance as shown in Figure 7b and c and cannot be transformed back to metallicity.

In order to obtain detailed information about oxygen exchange when measuring the resistance as function of oxygen activity, we analyze the amount of oxygen removed during a second reduction of oxidized BTO crystals. After oxidation at 800, 900, and 1000 °C, the sample is cooled to 350 °C under oxidizing conditions and subsequently reheated in UHV to the same temperature as used in the initial reduction step, while the oxygen effusion of atomic $^{16}O$ (Figure 7e) and molecular $^{16}O_2$ with mass 32 u (Figure 7f) is recorded by mass spectrometry. In contrast to the reduction process of the virgin crystal, in which the dominant outflow is atomic oxygen [22], there is a significant effusion of molecular $O_2$ from the oxidized sample and the total amount of effused oxygen is 1–2 orders of magnitude higher (Figure 7g). Hence, a paradoxical situation arises. Removing a small amount of oxygen from a stoichiometric crystal leads to an I/M transition, whereas a metallic state does not form when even more oxygen is effused during the second reduction of an oxidized crystal.

**The role of extended defects in channeling the electrical conductivity in oxidized crystals**

In our previous publication [22], we have examined the initial transformation of BTO crystals into a metallic state in detail and found evidence that electrical transport is channeled along the reduced cores of dislocations. Sayyadi-Shahraki et al. also reported on the increase of electrical conductivity in BTO by the introduction of dislocations [64]. Upon reduction, these extended defects can be switched into a metallic state due to a local agglomeration of oxygen vacancies, even at an extremely low global concentration of oxygen vacancies. As local conductivity effects related to conducting paths along dislocations play such a prominent role in reduced BTO, we investigate these effects for oxidized BTO. We analyze the surface of a BTO crystal, which was oxidized at 1000 °C by LCAFM. The topography map (Figure 8a) reveals complex topographic changes associated with recrystallization processes. The simultaneously recorded current map of the same area (Figure 8b) exhibits a significant heterogeneity in local conductivity. The overlay of topography and current map (Figure 8c) illustrates that there is a clear correlation between the position of the presumably dislocation-rich boundaries between the surface crystallites and the exits of conducting filaments or filament bundles indicating that the dislocations/filaments serve as diffusion pathways for not only oxygen, but also for Ba or BaO complexes. This allows for segregation and recrystallization processes during oxidation on the surface layer between dislocations at relatively moderate temperatures resulting in the formation of BaO-rich crystallites (Figure 3). The pronounced localization of current flow at crystallite boundaries also indicates that after oxidation, the filaments play a similar role to that in stoichiometric BTO crystals, namely, they can channel electrical transport. Hence, the matrix of the recrystallized regions continues to act as a “spectator” to current flow.

In order to investigate how such an inhomogeneously conducting surface reacts on electric polarization, we scan the oxidized surface while a high voltage was applied to the tip. Subsequently, we record current maps while the LCAFM tip is directly connected to a virtually grounded I/V converter, creating a short-circuit condition (Figure 8d). Thus, we can record the depolarization current over time. The current map is inhomogeneous, indicating that the filaments with an estimated density of $\sim 10^{10}/cm^2$ act as long-time sources of the depolarization current. The integration of the current reveals that a significant ionic charge of 0.1 C/cm² obtained from ten LCAFM mappings in short-circuit conditions at 120 °C is released (Figure 8e). This depolarization charge cannot be classified as a ferroelectric polarization, which typically lies below 0.1 mC/cm² [44]. Instead, this giant discharge effect indicates that ion accumulation and transport in the surface layer plays an important role. Polarizing the surface of oxidized crystals allows us to incorporate a large amount of ionic charges into the upper part of the crystal, i.e. an accumulation of ionic charges in volume of the surface region. When we perform the same polarization experiment on the cleaved surface of stoichiometric BTO we do not observe such a comparable depolarization current for the same low polarization voltage. As the accumulation of ionic charges in the surface layer at low temperatures and voltages is not expected to be related to bulk diffusion, we assume that ion transport occurs along the dislocation network in the surface-region. In this way, a transfer of oxygen ions between layers with different Ti oxidation states in oxidized BTO could be realized. The EMF of such a solid-state concentration cell (see Figure 5b) can induce a sufficiently high electrical field, leading to the self-polarization effect discussed above.

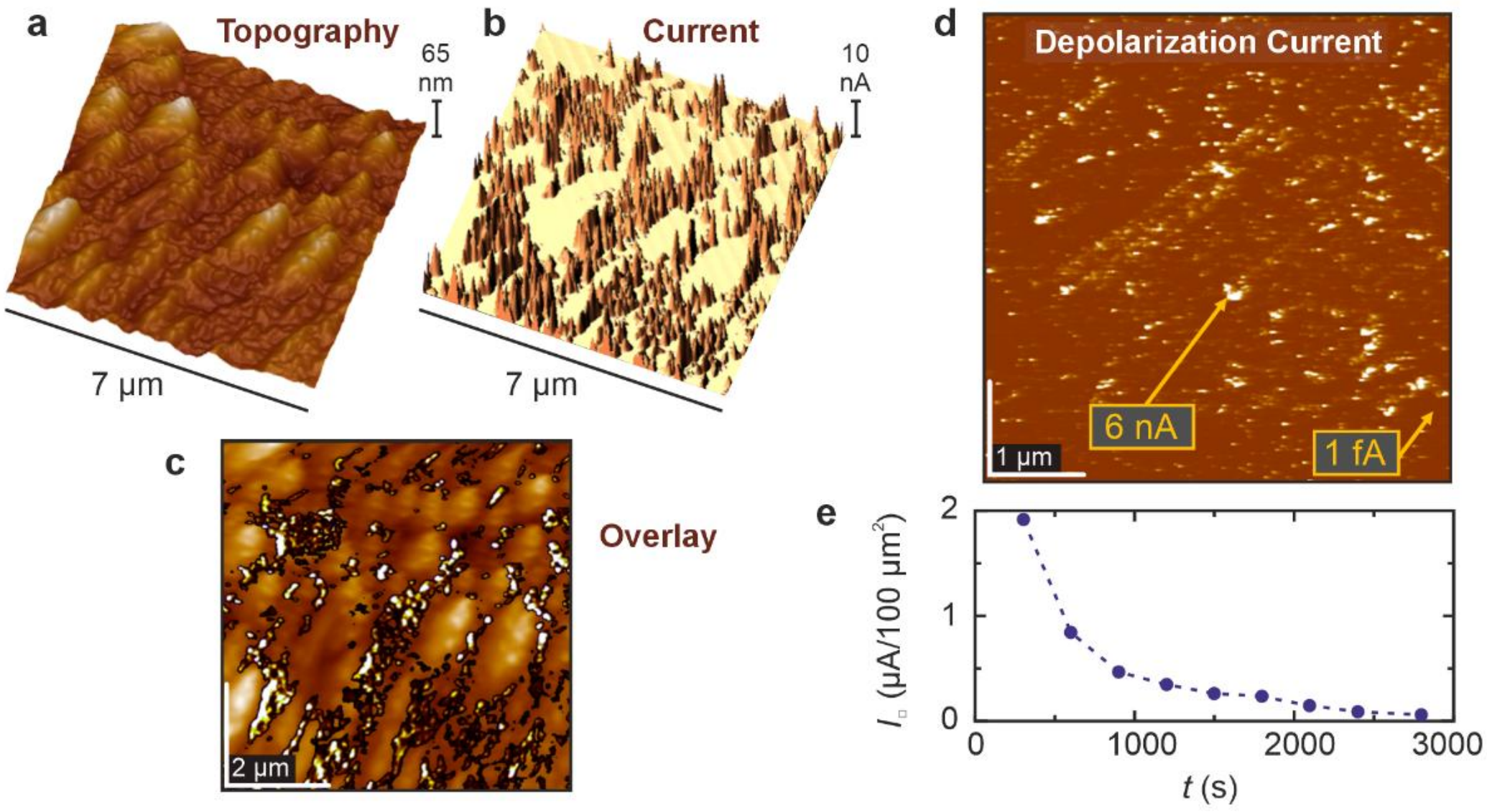

**Figure 8** LCAFM mapping of an oxidized BTO crystal. a) topography, b) current map, c) overlap of LCAFM and topography/current "grain analysis" (cut-off level/threshold 0.1 nA), d) LCAFM map of the depolarization current measured after polarizing the surface locally via the tip, e) depolarization current integrated over the mapping area of d) as a function of time.

In the following, we discuss the possibility of pipe diffusion in BTO in more detail. It can be expected that the massive recrystallization observed in oxidized BTO is associated with the easy diffusion of cations (here, Ba) along dislocation cores. While cation pipe diffusion in perovskites is widely accepted, and has been experimentally confirmed in BTO by isotope experiments [65], the possibility of oxygen pipe diffusion remains under debate [66–69]. To elucidate whether the easy diffusion of oxygen plays a role in BTO, we perform an $^{18}O$ reoxidation experiment on previously reduced BTO crystals and analyze the diffusion using SIMS depth profiling (Figure 9a). Note that in this case we investigate the non-equilibrium chemical diffusion and not tracer diffusion. For each reducing temperature the reduced crystals are exposed to 100 mbar $^{18}O_2$ for 30 minutes. This oxidation process represents the situation for determining the conductivity-oxygen activity diagram with our physical method. We observe that the normalized diffusion profiles of $^{18}O$ exhibit two distinct regimes for temperatures between 500 and 900 °C; a fast decay in the upper surface layer (for short SIMS sputtering time) and a plateau-like region in deeper parts of the crystal. This indicates that the chemical diffusion can be ascribed to a combination of bulk and pipe diffusion as in the Harrison "B" model [70]. Following this concept, we can estimate the activation energy of the easy diffusion process along extended defects/dislocations in the temperature range from 600 to 900 °C $E_a$ = 0.78(1) eV (Figure 9b). Considering that the presence of filaments in the surface region of oxidized BTO can facilitate ionic transport, one can conclude that an enormous accumulation of ions in the surface region can be easily achieved when the surface is externally polarized, even in the ferroelectric phase (see Figure 8d).

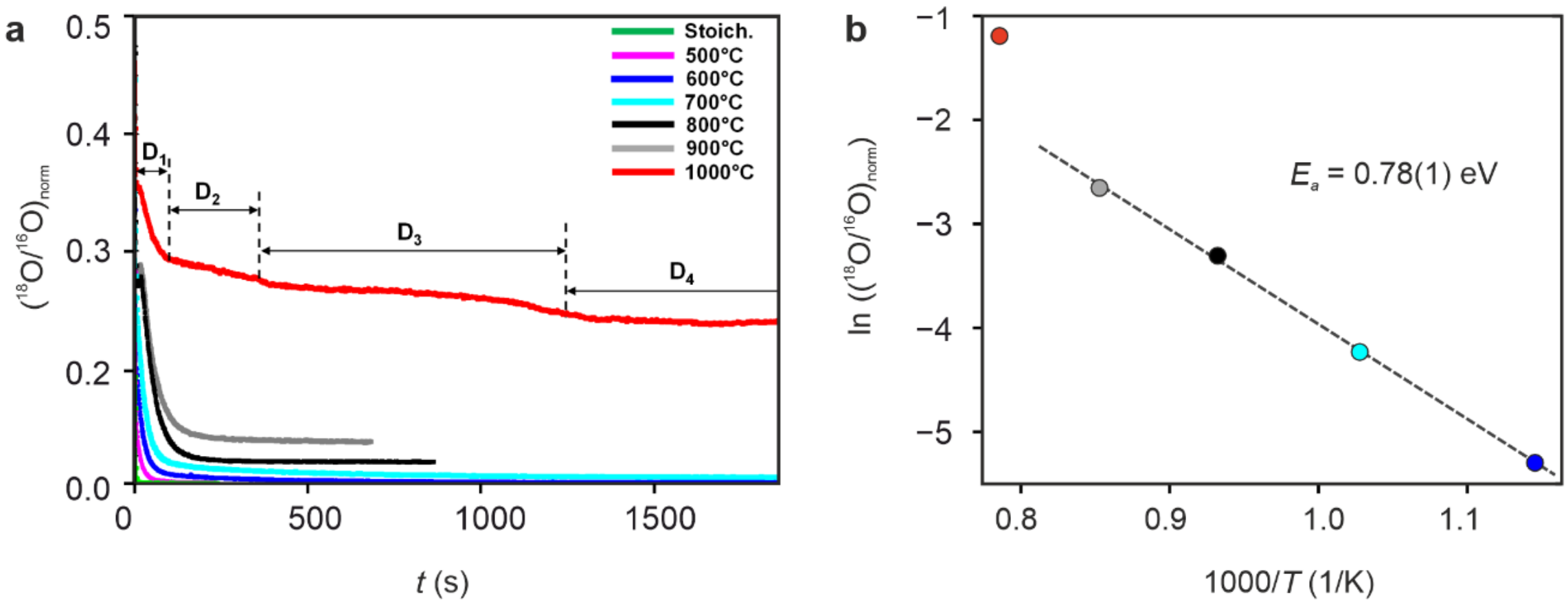

**Figure 9** Chemical oxygen isotope diffusion in reduced BTO. a) $^{18}O$ diffusion profiles obtained using TOF-SIMS analysis for BTO crystals reoxidized at different temperatures; b) Arrhenius plot of the normalized concentration of $^{18}O$ obtained from the plateau region of the diffusion profiles in a); for the calculation of the activation energy, only the data from 600–900 °C are considered.

In contrast to the diffusion profiles for lower reoxidation temperatures, which show only one plateau, the profile of the sample reoxidized at 1000 °C exhibits many fragments with parallel segments along the SIMS sputtering time axis, and different concentrations of the $^{18}O$ isotope can be identified in the profile (Figure 9a, $D_1$–$D_4$). To our knowledge, such an unusual depth profile has not been reported in the literature for ternary oxides with a perovskite structure and may reflect the influence of the hierarchical structure of the dislocation network and a significant change in the chemistry of the recrystallized surface region.

## IV Discussion and conclusion

Our results reveal that the oxidation of BTO, a model material for ferroelectrics and ternary oxides leads to significant segregation effects causing stoichiometric and structural transformations in the surface region, which directly influence the ferroelectric characteristics of the whole sample. Comparing our results with the comprehensive literature on ferroelectricity in perovskites, we realize that segregation effects in the surface region are seldom considered. There are studies discussing the role of the surface layer in relation to the influence of crystal's thickness on ferroelectricity [71], the tetragonality of the surface above $T_C$ [72], the residual pyroelectric effect above $T_C$ [73], the influence of the field and frequencies on the dielectric constant above $T_C$, and the nonlinear optical properties of an electrically polarized crystal above $T_C$. Most publications attribute these specific properties of the surface layer to a space-charge zone [74] or to chemically or mechanically distorted layers [75–77]. While the chemical influence of the surface layer on ferroelectric properties appears to align with our results, the finding that BTO can be turned into a metallic state upon thermal reduction by removing an extremely low amount of oxygen necessitates a different perspective. Despite the metallic behavior of the resistance, reduced BTO still exhibits ferroelectricity, thus behaving as a composite with both ferroelectric and metallic characteristics [78,79]. Hence, it should be regarded as a composite of metallic filaments formed at the core of dislocations [80] and the surrounding ferroelectric matrix [22]. As we have observed similar effects of surface segregation and the formation of metallic filaments in the surface layer upon initial reduction in further perovskite crystals including ferroelectric $PbTiO_3$ [50] and paraelectric $SrTiO_3$ [7], it appears that they are relevant for a variety of functional materials but are frequently neglected when building thermodynamic models.

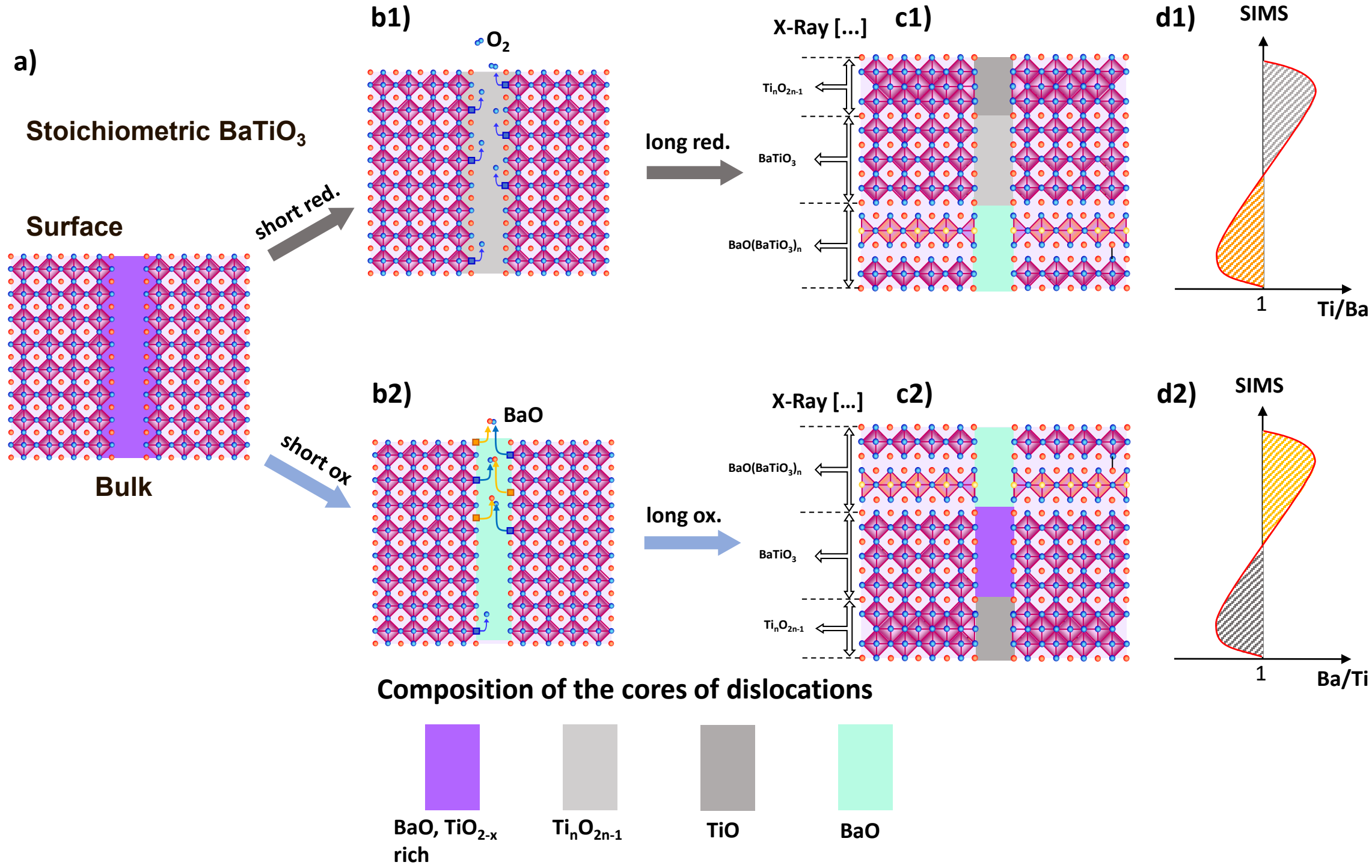

**Figure 10** Schematic illustration of the transformations of the dislocation-rich surface of BTO upon reduction and oxidation. a) stoichiometric surface of a "real" crystal, b) easy diffusion along dislocations, c) local segregation phenomena, d) trends in the change in the Ti/Ba (d1) and Ba/Ti (d2) ratios (outside the plane) during redox processes. Different colors mark the segments of dislocations with different chemical compositions.

Based on our nanoscopic results, we suggest a model describing the transformation of the dislocation cores upon reduction and oxidation as illustrated in Figure 10. Starting from a stoichiometric BTO surface containing dislocations (Figure 10a), a short-time reduction leads to the preferential removal of oxygen along the dislocation core, while a prolonged reduction results in the evolution of a Ti-rich surface composed of titanium suboxides such as Magnéli phases (Figure 10b1-d1). Similarly, we consider the high-temperature oxidation of BTO, which particularly affects the surface region and locally generates compositional inhomogeneity. Upon short oxidation, Ba cations or BaO complexes migrate along dislocations, resulting in Ba enrichment at the surface. As confirmed by various methods, including AFM, SIMS, XPS, and XRD, prolonged oxidation leads to recrystallization processes and phase transformations in the surface region. We conclude that this process involves the stacking of layers with different compositions in an opposite ordering to that observed during reduction (Figure 10b2–d2). Such a phase transformation at relatively low temperatures (500–1000 °C) is not expected from simple point-defect chemistry. Hence, in our opinion, the significant concentration of extended defects in the surface region and the easy transport of barium and oxygen along these defects must be considered to obtain a comprehensive description of the oxidation process. As a mechanism for the growth of new phases upon reduction and oxidation, we propose the so-called “zipper effect,” which describes the intercalation and deintercalation of BaO planes and is related to crystallographic shearing, as observed in STO [14]. Our model based on nanoscale phase transformations is supported by findings of Bursill et al., who have demonstrated that stoichiometric BTO structures can be converted to BaO- and $TiO_2$-rich compounds upon electron irradiation [81]. Emmez et al. have shown that low-temperature segregation of BaO on a $TiO_2$ layer can lead to the formation of BTO or R-P phases at 300 °C [82], suggesting that facile diffusion paths are relevant for these solid-state reactions. For in situ reduced BTO, our investigations have shown that stoichiometric changes in the surface layer of Ba can be observed by XPS below 500 °C [22].

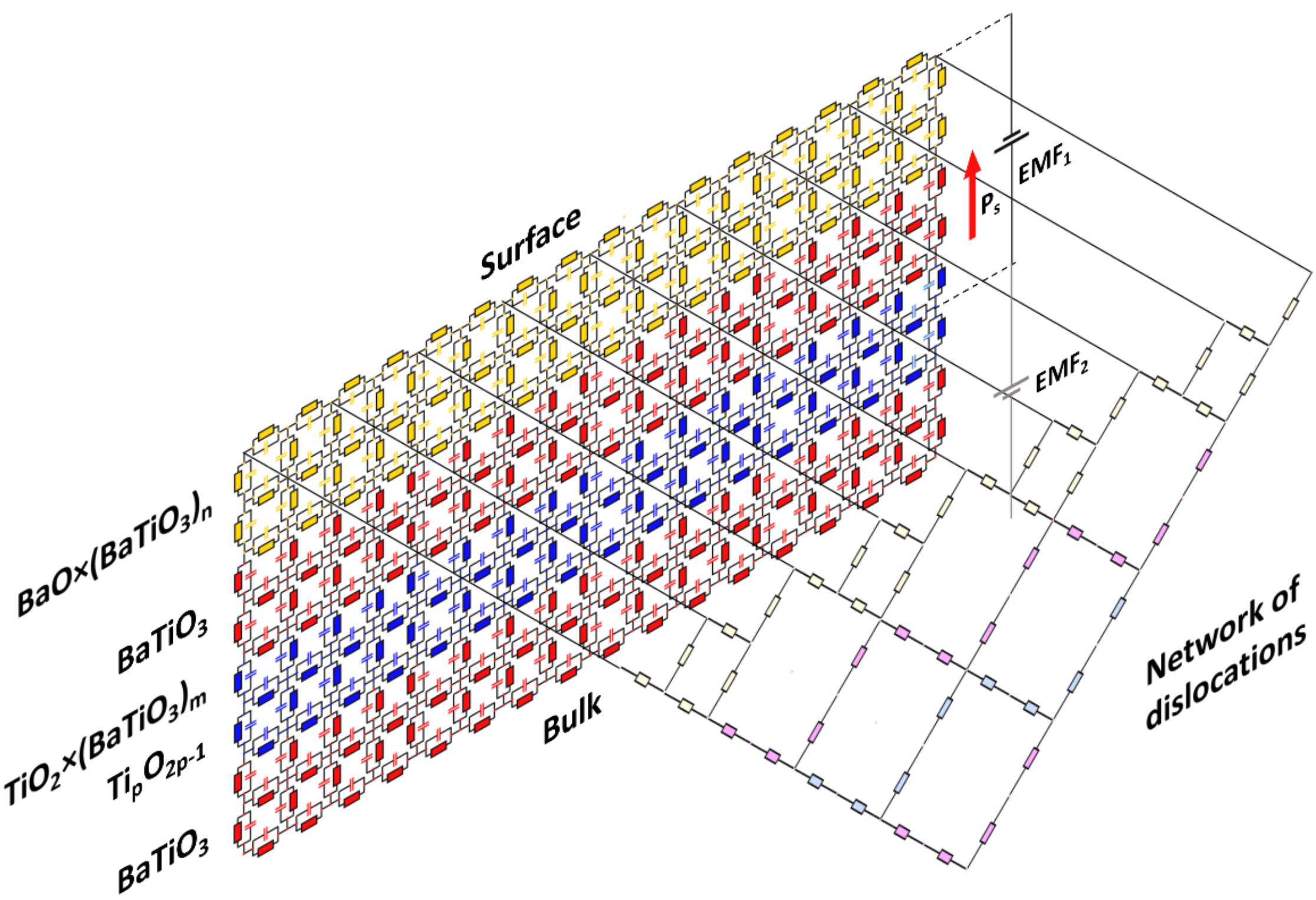

**Figure 11** Equivalent circuit of the surface region combining a network of capacitors and resistors representing the layers of different dielectric constants determined by the local chemistry and crystallography with a network of resistors representing the dislocation network connected in parallel.

We can expect that a new “chemical sandwich” structure is created in the surface region of a BTO crystal upon oxidation, which also determines the crystal’s electric behavior during the transition between the ferroelectric and paraelectric phases. While the details of the segregation processed on the free surface, which we access, e.g., by XPS and AFM, and below oxygen permeable top electrodes may slightly differ, the general tendencies in the segregation and transformation in BTO are the same as confirmed by XRD analysis on samples covered with thin electrodes see e.g., [36]. As shown in Figure 4, the capacitance curve of oxidized BTO does not exhibit the sharp first-order transition characteristic of stoichiometric crystals but instead becomes diffuse and shows many small peaks close to $T_C$. This illustrates the significant influence of the surface region with inhomogeneous segregation effects resulting in a local shift of $T_C$. Hence, the surface region cannot be directly associated with the traditional definition of a skin, as proposed by Känzig [74] and was solely based on space-charge and strain effects. To illustrate the influence of a sandwiched surface layer on capacitance, Figure 11 presents an equivalent circuit consisting of a network of resistors and capacitors that represent the stacked layers with different chemical compositions, namely $BaO\times(BaTiO_3)_n$ and $TiO_2\times(BaTiO_3)_m$. The dielectric response of such a Maxwell-Wagner capacitor would cause a slight modification to the $C(T)$ dependence of the virgin crystal because the bulk of the crystal still is in the original state. Additionally, we modelled the short-circuit effect of the dislocation network by connecting a resistor network with low resistance in parallel to the Maxwell-Wagner layers of the surface.

The dramatic reduction in polarization observed in the ferroelectric hysteresis loops upon oxidation (Figure 4) can be related to the imprint of internal electrical fields induced by the layer-like transformation of the surface region illustrated as $EMF_1$ and $EMF_2$ in Figure 11. The EMF is electrochemical in nature, similar to those found in oxidized $SrTiO_3$ [14] and $PbTiO_3$ [50]. Despite the low voltage of the EMF caused by the difference in reduction potential of the different layers, the potential drop occurs over a very short distance of a few dozen nm as confirmed by SIMS depth profiling. Hence, the local electrical field can exceed the coercive field, causing self-polarization. This is shown macroscopically for an oxidized crystal by removing one side of the surface region from the crystal, or nanoscopically using AFM potentiometry across the cross-section of an oxidized crystal [8]. Our data on multiple thermal depolarization and polarization cycles of this region without external polarization indicate that, due to the EMF in the chemically stacked layer, the surface region is self-poled (see Figure 5).

The creation of a complex layer structure in the surface region not only influences ferroelectric behavior but also affects further physical and chemical properties such as the electric conductivity. Our analysis of the electrical transport properties at varying oxygen activities reveals a lack of reproducibility across repeated oxidation and reduction cycles illustrating that continuous changes in the surface layer's structure and composition might prevent reaching a true equilibrium state in commonly employed experimental times. Only for oxygen activities near the intrinsic conductivity minimum of the Brouwer diagram, a reversible transformation between BaO-$BaTiO_3$ and stoichiometric $BaTiO_3$ has been observed at 1000 °C [33], while for highly reducing or oxidizing conditions irreversible segregation effects have to be taken into account. Hence, not only oxygen vacancies should be considered in the analysis of defect creation for temperatures between 500 and 1000 °C, but also chemical changes in the surface region. Furthermore, complex crystallographic shearing mechanisms must be considered to explain crystallographic phase transformations at moderate temperatures related to Magnéli or Ruddlesden-Popper phases. Such processes also require the easy migration of BaO complexes along dislocations, similar to the formation of SrO droplets at the exit of dislocations in oxidized STO, and the dismantling and intercalation of the BaO plane and the generation of hairpin dislocations, which are connected with crystallographic shearing for the creation of the Magnéli phases [83]. In our opinion, the preferential generation of oxygen vacancies at dislocations and the easy diffusion along the 3D dislocation network in the surface region can be made responsible for the effect of switching a virgin crystal to a metallic state upon reduction under vacuum conditions at an extremely low global oxygen vacancy concentration. However, after subsequent oxidation, the same reduction procedure does not restore metallic conductivity, which shows that a one-time-oxidized crystal has a different chemical, crystallographic, and electronic structure in the surface region than the original stoichiometric crystal. Due to the properties of the surface region, the two are different objects, although the original crystal "remains hidden" in the bulk. The differences concern not only the chemical composition but also the fact that dislocations pass through layers of different composition. Therefore, they cannot be analyzed in the same way as those of the crystal reduced for the first time, in which all segments of the dislocation core are metallic. After oxidation and repeated reduction, however, the dislocation core segments are not metallic in regions with a surplus of BaO (e.g., R-P phases). Hence, these semiconducting fragments of the filaments, which are galvanically connected to the metallic TiO segments of the TiO-rich regions, suppress the formation of global metallicity across the entire network of dislocations.

Given the complex interplay of processes that accompanies oxidation and reduction, we acknowledge that our interpretation is not comprehensive and requires further research and in-depth analyses. However, our results clearly indicate that the transformation of crystalline BTO during reduction and oxidation is complex. Considering that numerous articles propose that in polycrystalline BTO at high temperatures ($T > 1000$ °C), the thermodynamic description based on the statistically distributed point defects can be used, our results suggest that this picture should be adapted and expanded in order to understand the complex mechanisms upon reduction and oxidation of BTO. This will be particularly relevant when processing BTO at lower temperatures ($T < 1000$ °C), which favor pipe diffusion over bulk diffusion thus resulting in inhomogeneous segregation phenomena intrinsically.

**Acknowledgements**

We are indebted to Clive Randall for valuable discussions.

**Data availability**

The data that support the findings of this article are not publicly available. The data are available from the authors upon reasonable request.